# Chapter 4

# RF Systems

*P. Baudrenghien[1], G. Burt[2], R. Calaga[1*], O. Capatina[1], W. Hofle[1], E. Jensen[1], A. Macpherson[1], E. Montesinos[1], A. Ratti[3] and E. Shaposhnikova[1]*

[1]CERN, Accelerator & Technology Sector, Geneva, Switzerland
[2]University of Lancaster, Lancaster, UK and Cockcroft Institute Sci-Tech Daresbury, Warrington, UK
[3]FNAL, Fermi National Accelerator Laboratory, Batavia, USA

## 4     RF systems

### 4.1    Introduction

The HL-LHC upgrade to enhance the integrated luminosity by a factor of 10 per year will need the following RF systems:

- deflecting (or crab) cavities for compensation of the effective geometric crossing angle at the IP to recover the luminosity loss due to increased crossing angle;
- a harmonic RF system for bunch manipulation and increased stability;
- a transverse damper upgrade for higher power, bandwidth, and low noise.

The above RF systems are described with relevant technical details below. The beam and machine parameters from Appendix A.1 are used to design the RF systems.

### 4.2    Crab cavities

The LHC uses a 60 m common focusing channel plus 21 m common drift space and 20 m common dipole channel on each side of the interaction region (IR), where the two counter-rotating beams have to be separated transversely to avoid parasitic collisions. Separation is accomplished by introducing a crossing angle at the interaction point (IP), which needs to increase with the inverse of the transverse beam size at the collision point in order to maintain a constant normalized beam separation. The non-zero crossing angle implies an inefficient overlap of the colliding bunches. The luminosity reduction compared to that of a zero crossing angle, assuming a Gaussian distribution, can be conveniently expressed by a reduction factor,

$$R_\Phi = \frac{1}{\sqrt{1+\Phi^2}} \;, \tag{4-1}$$

where $\Phi = \sigma_z \phi / \sigma_x$ is the aspect ratio of the longitudinal ($\sigma_z$) to the transverse ($\sigma_x$) beam sizes multiplied by the half crossing angle $\phi$, which is also known as the Piwinski angle [1]. Alternatively the reduction can be viewed as an increase in the transverse beam size at the collision point to effective beam size given by $\sigma_{\text{eff}} = \sqrt{\sigma_x^2 + \sigma_z^2 \phi^2}$. For HL-LHC beam parameters, the reduction compared to the case of a head-on collision can be 70% or larger. Therefore, the effective gain in luminosity by simply reducing the beam size at the collision point diminishes rapidly.

   To recover the loss, an elegant scheme using RF deflectors (also known as crab cavities) on either side of the collision point was first proposed and used for electrons [2, 3]. The time-dependent transverse kick from an RF deflecting cavity is used to perform a bunch rotation, in the *x–z* plane or *y–z* plane depending on the crossing angle orientation, about the barycentre of the bunch (see Figure 4-1). The kick is transformed to a relative displacement of the head and the tail of the bunch at the IP to impose a head-on collision while maintaining the required beam separation to minimize parasitic collisions. The upstream RF deflector is used to reverse the kick to confine the bunch rotation to within the IR. The crab crossing scheme in a global compensation using only a single cavity per beam was successfully implemented at the e⁺e⁻ collider at KEKB in Japan to achieve record luminosity performance [4].

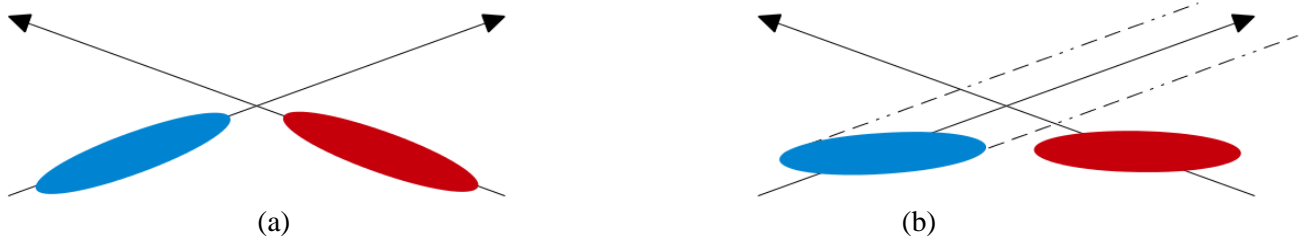

Figure 4-1: Bunches colliding with (a) a crossing angle without crab crossing; (b) with the crab crossing

Since the luminosity gain is substantial, the crab crossing scheme is adopted as a baseline for the HL-LHC upgrade. The time-dependent transverse kick can equally be used to regulate the crossing angle at the IP and therefore allows for a natural knob to control the total number of events per crossing (luminosity levelling), a feature highly desired by the experiments. Levelling by means of collision offsets is already used at LHCb and ALICE. More sophisticated means of levelling to control the density of the events along the luminous region by means of crab cavities are under investigation [5].

4.2.1 Beam and RF system parameters

The HL-LHC upgrade lattice requires crab cavities to provide a total voltage of 12–13 MV per beam per side of each collision point at a frequency of 400.79 MHz. Since the crossing plane in the two experiments is different, a local crab cavity system is a prerequisite. The nominal configuration will use a two-cavity cryomodule as the basic unit. Four such modules per IP side at P1 and P5 are required to perform the bunch rotation resulting in 16 modules (a total of 32 cavities). Four spare modules (eight cavities) are required. Two are designed for horizontal crossing and the other two for vertical crossing to perform crab crossing for P5 and P1, respectively. The low frequency is required to minimize the RF curvature for the long LHC bunches (see Appendix A.1). The machine constraints near the interaction region require cavities with a transverse dimension compatible with the location of the adjacent beam pipe. The RF and machine parameters directly relevant to the crab cavities are shown in Table 4-1.

An operating temperature of 2 K is chosen as a baseline. A pressure stability on the cavity surface should be minimized to less than 1 mbar. The static and dynamic heat load is expected to be approximately 30 W to the 2 K bath for a two-cavity module. A cavity vacuum level to better than $10^{-10}$ mbar is required to assure stable performance.

The input RF power of 80 kW is available to power each of the eight cavities to their nominal voltage with sufficient margin to cope with beam loading caused by beam offset. The low level RF (LLRF) will include a regulation loop around the tetrode (to reduce the RF amplitude and phase noise in a band extending to a few tens of kHz), plus an RF feedback to control the vector sum precisely on the two sides of the interaction region. Longitudinal pickups (Pus) located close to the crab cavities (one per IP-side per beam) are used to regulate the slow drifts of the deflecting voltage with respect to the average bunch centre. To stay within the specified RF power limits, an orbit stability including mechanical tolerances must not exceed 1 mm with stable beams at flat-top. The cavity is kept on tune at all times. The resonant frequency should be precisely controlled by a tuning system to a level well below 0.5 kHz to be compatible with the RF power limits.

Table 4-1: Relevant RF parameters for HL-LHC crab cavities

| Characteristics | Units | Value |
|---|---|---|
| Resonance frequency | [MHz] | 400.79 |
| Bunch length | [ns] | 1.0 (4 $\sigma$) |
| Maximum cavity radius | [mm] | ≤145 |
| Nominal kick voltage | [MV] | 3.4 |
| $R/Q$ (assumed, linac convention) | [Ω] | 400 |
| $Q_0$ |  | ≥1 × $10^{10}$ |
| $Q_{ext}$ (fixed coupling range) |  | 3 × $10^5$–5 × $10^5$ |
| RF power | [kW] | 80 |
| Power coupler OD (50Ω) | [mm] | 62 |
| LLRF loop delay | [μs] | ≈1 |
| Cavity detuning | [kHz] | ≈1.0 |

4.2.2 RF cavity design

In order to sustain the surface fields at the required kick gradient of 3.4 MV/cavity for the LHC, crab crossing superconducting technology is essential; space restrictions, voltage requirements, and impedance considerations strongly rule out a normally conducting option. 'Conventional' superconducting elliptical cavities, which have already been used at KEK, pose significant integration problems at the operating frequency of 400 MHz in the LHC due to their transverse size.

This led to the concept of 'compact' cavities. These cavities have unconventional geometries not widely used in superconducting technology. A few concepts with complex shapes exist primarily in the field of heavy ion acceleration. Such structures fit within the LHC constraints in the existing tunnel and reveal significantly better surface field characteristics than the conventional cavities for beam deflection. As a result of an intense R&D programme within the EuCARD and LARP programmes and with other external collaborators during the past four years, three compact designs at 400 MHz have emerged as potential candidates. Their topologies are shown in Figure 4-2. The three proposed designs are at least four times smaller in the plane of crossing compared to an elliptical cavity with a ratio of the kick gradient to the peak surface fields lower by a factor of 2.

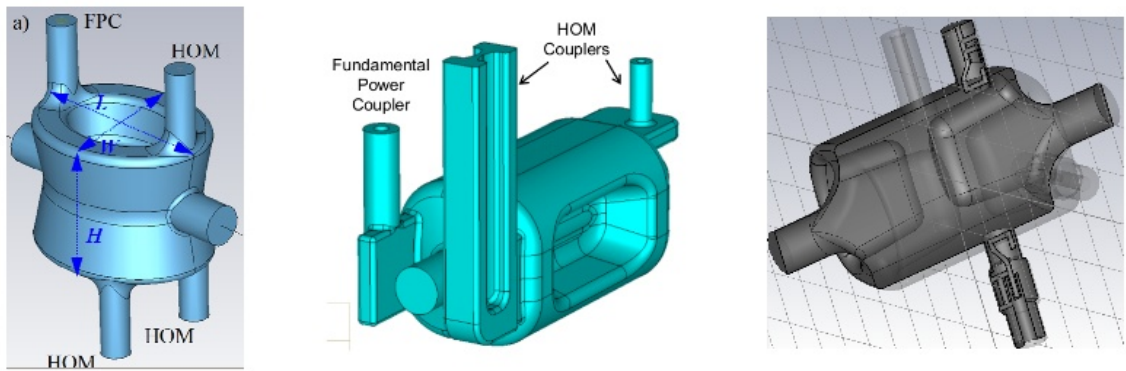

Figure 4-2: Compact cavities. (a): Double quarter wave cavity (DQW), courtesy of Brookhaven National Lab. (b) RF dipole cavity (RFD), courtesy of Old Dominion University. (c) Four-rod cavity, courtesy of Lancaster University.

As a part of the R&D phase, it was decided to prototype full-scale cavities for all three designs for a field validation at the nominal kick voltage. Following the recommendation by the Crab Cavity Advisory Panel a prototype module in a two-cavity configuration will be tested with beam in the SPS machine with LHC type beams [6]. These tests will help validate the cavity performance and operation with beam and understand the effects on protons as well as relevant machine protection aspects. The three cavities were fabricated in 2012–2013 and their performance validated at or beyond the nominal kick voltage [7–9]. The cavity designs including the fundamental power coupler and higher order mode couplers have evolved significantly from the prototype to meet the impedance requirements of the LHC. Following the recommendation of the May 2014 technical review [10], only two of the cavity designs have been considered for SPS testing (DQW and RFD).

The development of a two-cavity cryomodule for the SPS tests in 2017 is at an advanced stage. An overview of crab cavity planning spanning approximately ten years until full installation in the LHC is shown in Table 4-2. A more detailed plan for the SPS and the LHC, including the pre-series and series production, can be found in Ref. [35].

Table 4-2: Overview of crab cavity planning from R&D to installation in the LHC

| 2013–2014 | 2015–2016 | 2017–2018 | 2019–2023 | 2023–2024 |
|---|---|---|---|---|
| Cavity testing and prototype cryomodule | SPS cryomodule fabrication | SPS tests and LHC pre-series module | LHC cryomodule construction and testing | LHC installation |

4.2.3 Beam loading and RF power

In deflecting cavities operated in the crabbing mode, the RF phase and the RF component of the beam current are in quadrature ($\phi_s = 0$, synchrotron convention). For a beam transversely centred, there is no beam loading: the RF generator does not pass power to the beam. With a superconducting cavity (negligible surface losses) the RF power required to maintain the cavity voltage decreases monotonically with $Q_L$. Therefore, with a perfectly centred beam, the choice of $Q_L$ only requires sufficient bandwidth for unavoidable frequency transients due to external perturbations (see Section 4.2.10.4, Frequency tuning).

The situation is different for a beam circulating at an offset $\Delta x$. The beam-induced voltage due to an orbit offset is given by

$$\Delta V = I_b \cdot \frac{R_T}{Q_0} \cdot Q_L \cdot \Delta x, \qquad (4\text{-}2)$$

where $I_b$ is the average beam current, $R_T/Q_0$, is the transverse shunt impedance in Ω and $\Delta x$ is the offset. A sufficient bandwidth and the corresponding RF power are required to compensate for the unavoidable orbit offsets. Figure 4-3 shows the required forward power as a function of the $Q_L$ for a beam that is centred (red) and off-centred by 1 mm (green) and 2 mm (blue). It is expected that the orbit will be kept within 0.5 mm for the entire energy cycle of the LHC; another 0.5 mm should be added for mechanical tolerances.

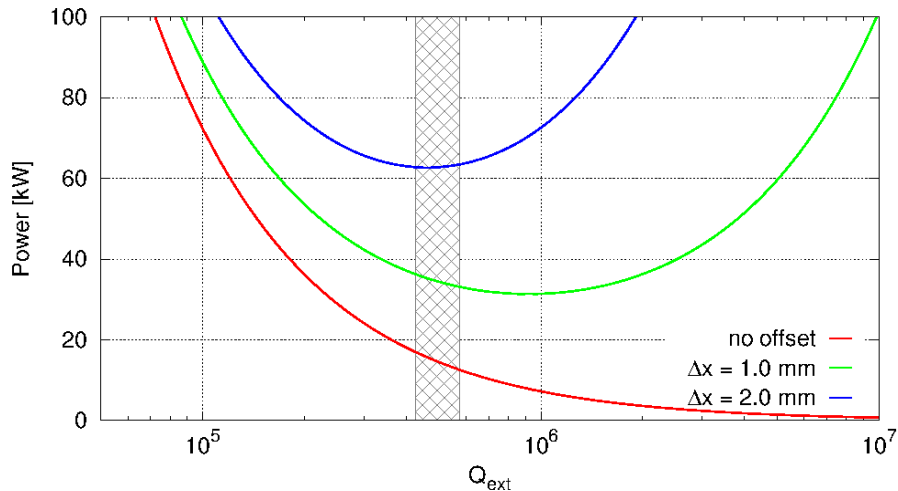

Figure 4-3: Forward power vs. cavity $Q_L$ for centred (red), 1 mm offset (green), and 2 mm offset (blue) beams. Assumed $R/Q = 400$ Ω, 3.4 MV RF, 1.1 A DC.

The required RF power has a broad minimum (≤ 40 kW) from a $Q_L$ of about $3 \times 10^5$ to $1.5 \times 10^6$ $1.5 \cdot 10^6$ for an offset specification of 1 mm. Selection of an optimal $Q_L$ value in the broad minimum is a compromise between the feasible tuning precision and the minimization of the field fluctuations from the amplifier electronics. For larger bandwidth (leading to more stability), lower $Q_L$ values are favoured – the cross-hatched area in Figure 4-3 was chosen as a compromise. A lower $Q_L$ is also favourable for the tuning system as it relaxes the precision needed by a mechanical system and the power needed to compensate for fast frequency changes.

### 4.2.4 RF power coupler

The RF power coupler was designed in view of the HL-LHC requirements; additional constraints (common platform) were introduced to limit the variances between the alternative designs in view of the SPS tests.

The crab cavity power coupler adopted will use a single coaxial disk-type window to separate the cavity vacuum and the air side. The antenna shape is specific to each cavity as the coupling mechanisms for the different cavities are not identical. However, a common platform starting from the cavity flange followed by the ceramic and double wall tube is imposed. To respect the common platform, the inner antenna is 27 mm diameter with an outer coaxial line of 62 mm diameter for a maximum power capability of approximately 200 kW. The inner line is made of a copper tube and the outer line is 316LN stainless steel with the inner surface coated with copper. The vacuum-to-air separation is achieved with a coaxial ceramic window ($Al_2O_3$) with an outer flange made of titanium. The rest of the items are built from massive Oxygen Free Electronic (OFE) 3D forged copper blocks. The coupler body is made in a conical line to increase the ceramic region to limit arcing, with the primary aim of enlarging the air side to the maximum while keeping the 62 mm/27 mm dimensions for the input antenna on the vacuum side. A coaxial to waveguide transition is performed with a WR2300 half-height without a doorknob (see Figure 4-4 (a)).

The air side of the coupler will be air-cooled while the antenna itself will be water-cooled. The waveguide design includes the possibility of DC polarization in order to avoid multipacting effects. Each coupler is equipped with three ports for a vacuum gauge, electron monitoring, and arc detection devices. The vacuum gauge, which is mandatory to protect the window during conditioning as well as in operation, will be oriented along the air line in order to minimize the cryomodule flange size. Special test boxes to condition the couplers have also been designed (see Figure 4-4 (b)). The coupler ports are designed to come out on the top of the cryomodule, perpendicular to the beam axis for ease of integration with the WR2300 waveguide transition. The cavity's helium vessel is designed to withstand the weight of the couplers and the waveguide (approximately 35 kg). The alternating crossing angle scheme will require that the orientation of a coupler assembly be robust for horizontal and vertical deflections.

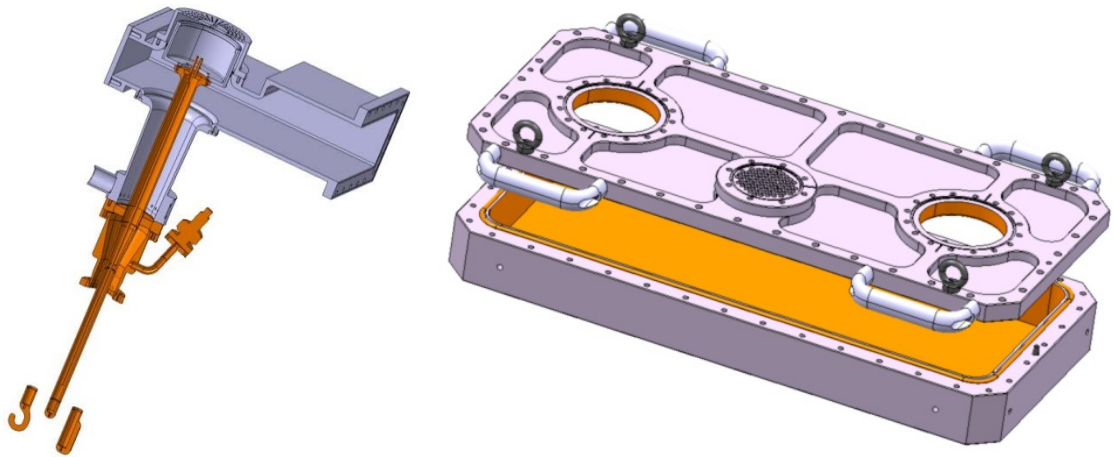

Figure 4-4: (a) Input coupler assembly; (b) test box for RF conditioning

### 4.2.5 Coupled bunch instabilities

The crab cavities must cope with the various modes of the collider cycle: filling, ramping, and physics. During filling of the 2808 bunches into the LHC, ramping, or operation without crab cavities (crabbing off), the cavity

85

can be detuned; but a small field should be kept for the active tuning system. This is referred to as 'parking'. Parking the cavity half the distance between two revolution frequency sidebands would be ideal for stability. Another possibility is to operate with 'crabbing off', which is possible since more than one cavity is used, namely counter-phasing to make the effective kick voltage zero while always keeping accurate control of the cavity field. This counter-phasing ensures both zero effective voltage and beam stability on tune – in fact, it has been found that this is the preferred scenario [11].

If detuning is used with a positive non-integer tune ($Q_h = 64.3$), the cavity should be tuned above the RF frequency to make the mode $l = -64$ stabilizing (see Ref. [11]). Although RF feedback is not mandatory for stability with a detuned cavity, it is preferred for accurate knowledge about, and control of, the cavity's resonance frequency and field. Active feedback will also keep the beam-induced voltage zero if the beam is off-centred. The additional RF power is used as a measurement of beam loading to guide beam centring. The RF signal picked up through the HOM couplers might also be used.

On the flat-top detuning can be reduced (but keeping the total kick voltage initially at zero). The RF feedback keeps the cavity impedance small (beam stability) and compensates for beam loading as the cavity moves to resonance. Once the cavity detuning is reduced to zero, we drive counter-phasing to zero and use the functions to synchronously change the voltage in all crab cavities as desired (crabbing on). In a physics run, with crabbing on, the active RF feedback will continue to provide precise control of the cavity field. The RF feedback reduces the peak cavity impedance and transforms the high $Q$ resonator to an effective impedance that covers several revolution frequency lines. The actual cavity tune has no big importance for stability anymore. The growth rates and damping rates are much reduced, and we have no more dominant mode as shown in Figure 4-5.

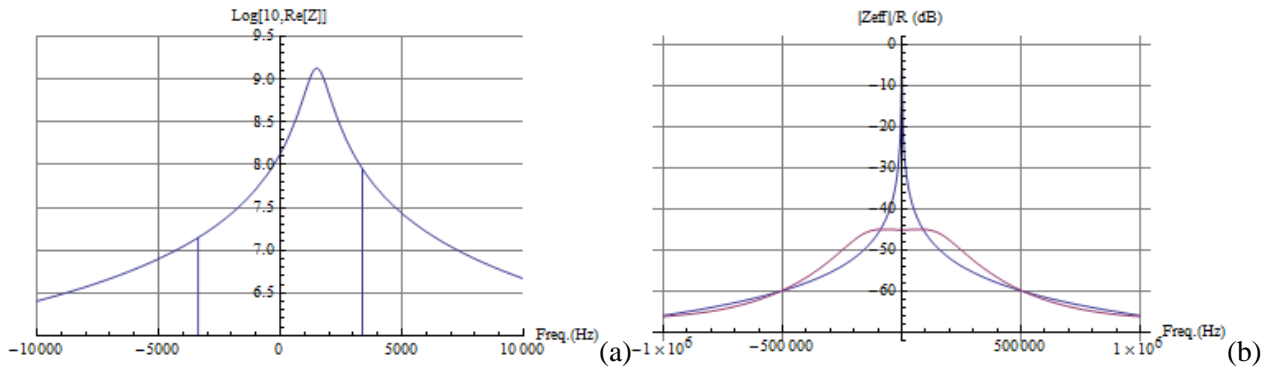

Figure 4-5: (a) Real part of the deflecting mode impedance with a detuning of 1.5 kHz from from 400 MHz. The vertical lines represent the difference in $\Re\{Z\}$ evaluated at $\pm 0.3$ $f_{\text{rev}}$ for the computation of damping rate (mode l = −64). (b) Modulus of the cavity impedance seen by the beam with the RF feedback on (red) and off (blue) normalized to the cavity impedance at the fundamental mode.

4.2.6 Impedance budget

On resonance, the large impedance of the fundamental deflecting (dipole) mode is cancelled between the positive and negative sideband frequencies, which are symmetric around $\omega_{\text{RF}}$. The active feedback will reduce the growth rates by a large factor.

For higher order modes (HOMs), both narrowband and broadband impedance should be minimized during the entire machine cycle as the LHC will accelerate and store beams of currents exceeding 1.1 A (DC). Tolerances are set from impedance thresholds estimated from Ref. [12].

The longitudinal impedance has approximately a quadratic behaviour vs. $f$ in the region of interest with the minimum threshold value at 300–600 MHz The total maximum allowed impedance from each HOM, summing over all cavities in one beam, assuming that the HOM falls exactly on a beam harmonic, is set at <200 kΩ, so if all 16 cavities have identical HOM frequencies, the longitudinal impedance must not exceed 12.5 kΩ per cavity. For frequencies higher than 600 MHz, the threshold is higher ($\propto f^{5/3}$), but the same threshold was imposed. Modes with frequencies above 2 GHz are expected to be Landau-damped due to natural frequency spread and synchrotron oscillations.

In the transverse plane, the impedance threshold is set by the bunch-by-bunch feedback system with a damping time of $\tau_D = 5$ ms [12]. Assuming the pessimistic case that the HOM frequency coincides with the beam harmonic, the maximum impedance is set to be <4.8 MΩ/m. Again, assuming 16 cavities per beam, the maximum allowed impedance per cavity is 0.3 MΩ/m. Analogous to the longitudinal modes, frequencies above 2 GHz are expected to be Landau-damped due to natural frequency spread, chromaticity, and Landau octupoles. It should be noted that there are nominally only eight cavities per transverse plane, so the threshold per cavity is higher, but 0.3 MΩ/m is given assuming that the crossing plane between the experiments could become the same as a worst case scenario.

Due to the very tight impedance thresholds, the distribution of HOM frequencies due to manufacturing errors can help relax the tolerances. The beam power deposited in the longitudinal HOMs can become significant when the frequencies coincide with bunch harmonics. The HOM couplers were dimensioned to accept a maximum of 1 kW to be able to cope with HL-LHC beams [13].

### 4.2.7 Higher order mode couplers

The first design goal of the HOM filter is to block the transmission of the main deflecting mode, while transmitting all remaining HOMs. Several HOM coupler designs were developed and optimized for different cavity geometries. Two high-pass filter designs, incorporating a notch filter at the fundamental frequency, are shown in Figure 4-6 with both HOMs using hook-like antennae to couple to the HOMs.

Simulations show that the HOM coupler must have a superconductive surface due to the high fields of the fundamental mode. A second design constraint requires that HOM couplers be able to efficiently remove the power in the HOMs (up to 1 kW) and the heat dissipated by the fundamental mode in the inner part of the HOM coupler from the cavity. High purity bulk niobium with sufficient cooling can ensure this. The required cooling may be possible by conduction, but the possibility of actively cooling with superfluid liquid helium or immersion in a small He tank is also under study.

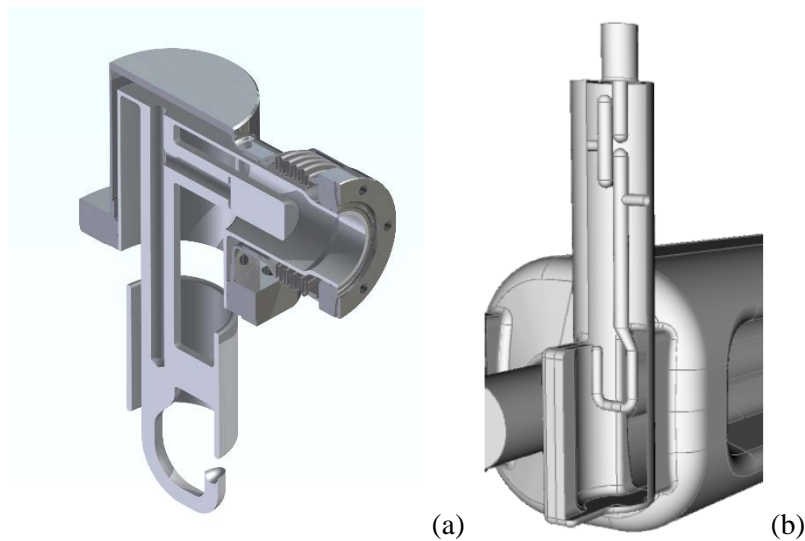

(a)      (b)

Figure 4-6: HOM filter for (a) DQW; (b) RFD

### 4.2.8 RF multipoles

The crab cavity designs presently considered are such that they lack axial symmetry. Therefore, they can potentially exhibit all higher order components of the main deflecting field. Due to the placement of the cavities at high beta-function locations, the higher order components of the main deflecting mode can affect long-term particle stability. RF multipole components $b_n$ of the RF deflecting field can be approximated and hence expressed in a similar fashion to magnets:

$$b_n = \int_0^L \frac{1}{qc} F_\perp^n \, dz \quad [\text{T m}^{2-n}] . \tag{4.1}$$

The quadrupolar component $b_2$ is zero in the case of perfect symmetry; due to fabrication errors and ancillary components it is non-zero – it must be smaller than 10 units leading to a tune shift in the order of $\Delta Q \approx 10^{-4}$. The first systematic multipole is the sextupolar component, $b_3$. Long-term simulations with the optical functions of the HL-LHC indicate that the $b_3$ component should be limited to approximately 1000 ±10% units, which results in an acceptable degradation of the dynamic aperture below $1\,\sigma$ for orbit offsets of 1.5 mm [1]. No specifications are yet provided for higher order terms, but it is expected that they be controlled to smaller values than the neighbouring D2 dipole magnet.

For $n \geq 4$, assuming a very approximate scaling of the additional kick from an orbit offset via $b_n$, the $b_n$ must be kept $< O(10^n)$. Better estimates are pending; results from long-term tracking are needed.

### 4.2.9 Lorentz force detuning and multipacting

When the cavity contains RF fields there is a Lorentz force on the cavity surface resulting from the high radiation pressure on the cavity walls. This results in a detuning of the cavity frequency. The Lorentz force detuning is kept small ($\leq 0.6$ kHz) at the nominal field.

Another common problem in complex RF structures is multipactor. This is a resonant electron phenomenon where an electron follows a regular trajectory in the RF fields, where it strikes the surface with energy such that the number of secondaries produced is statistically likely to be greater than one. If these secondaries follow the same trajectory then the process will repeat causing an exponential growth in the number of secondaries. The electrons will absorb RF power, limiting the field to a finite level and depositing additional heat load in the walls. Initially the cavity surface may have an oxidized layer that will increase the secondary emission yield (SEY). However, multipactor also conditions the surface, removing this layer. If the multipactor disappears after processing or if sufficient power is available to overcome the multipactor it is termed a 'soft' barrier; otherwise it is termed a 'hard' barrier.

Multipactor was modelled in all cavities and couplers using two codes using different methodologies to identify multipactor. CST Particle Studio uses particle tracking with accurate secondary emission models to simulate the growth in electrons with time, while Track3P tracks a single particle in the RF fields and looks for resonant trajectories.

In CST three SEY models were used to look at the effect of surface cleanliness. The models were for wet treated, baked, and processed niobium surfaces. While multipactor in all cavities was found for the wet-treated and baked models, no multipacting trajectories were found for the processed surface, suggesting that any multipactor would be soft and easily processed through. Similarly, Track3P found multipactor at low field. This is in good agreement with the results from the prototype tests, where multipactor was observed and could be processed away easily.

### 4.2.10 Cryomodule and integration

#### 4.2.10.1 Temperature choice

The BCS resistance of niobium at 4.5 K and 400 MHz is around 50 n$\Omega$, which is more than 10 times larger than the value at 2 K. The complex shapes of the cavities may also be susceptible to microphonics caused by liquid He boil-off, hence operation below the lambda point of He is preferred. For these reasons operation at 2 K is baseline. This will require the provision of liquid He at 2 K to the crab cavity location in the LHC. The current heat load limits for the LHC are not currently known, but are likely to be around 3 W of dynamic load per cavity at 2 K.

#### 4.2.10.2 Cavity interfaces and cold mass

Following the recommendation of the May 2014 technical review [10], only two cavity designs are considered for the engineering design towards the SPS tests (DQW and RFD). The mechanical design of the cavities ensures their safe use under maximum loading condition during their entire life-cycle. The cavity was dimensioned to cope with several mechanical constraints: ensure elastic deformation during maximum pressure as well as during all transport and handling conditions; maximize tuning range; minimize sensitivity to pressure fluctuation; avoid buckling due to external pressure; and maximize the frequency of the first mechanical natural mode. The final mechanical design of the cavities including all external interfaces is shown in Figure 4-7.

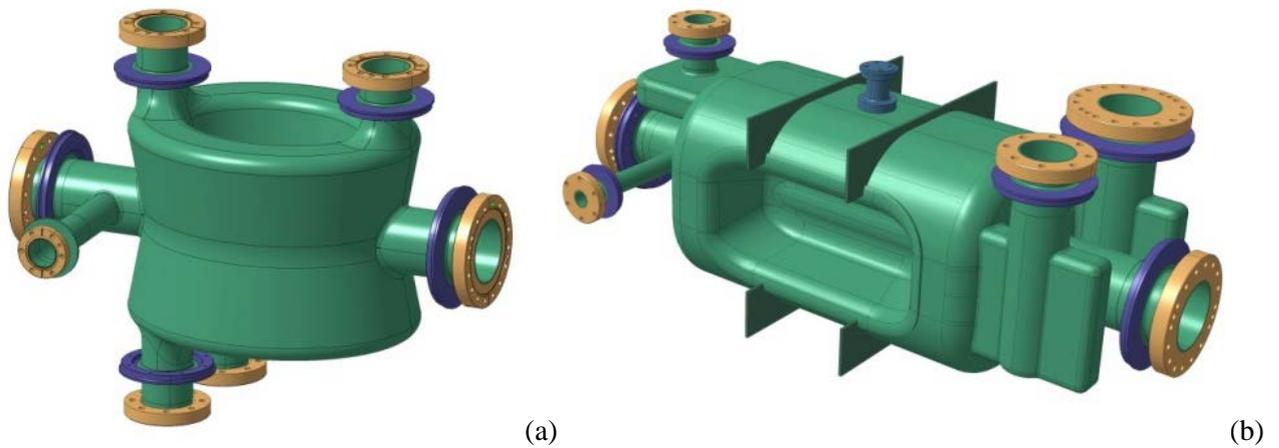

(a) (b)

Figure 4-7: Schematic view of the cavity with interfaces (a) DQW; (b) RFD

The superconducting resonators are fabricated from bulk niobium sheets by electron-beam welding of deep-drawn parts. A final thickness of 4 mm was calculated to be acceptable in order to cope with all the mechanical constraints as well as minimizing the cost of cavity production. The cavities are bath-cooled by saturated superfluid helium at 2 K. Each cavity is equipped with: a helium tank, a tuning system, a fundamental RF power coupler, a field probe, and two or three HOM couplers. A functional specification including all tolerances for the cavity with its interfaces to develop manufacturing drawings for the DQW and the RFD are shown in Figure 4-8 and Figure 4-9, respectively.

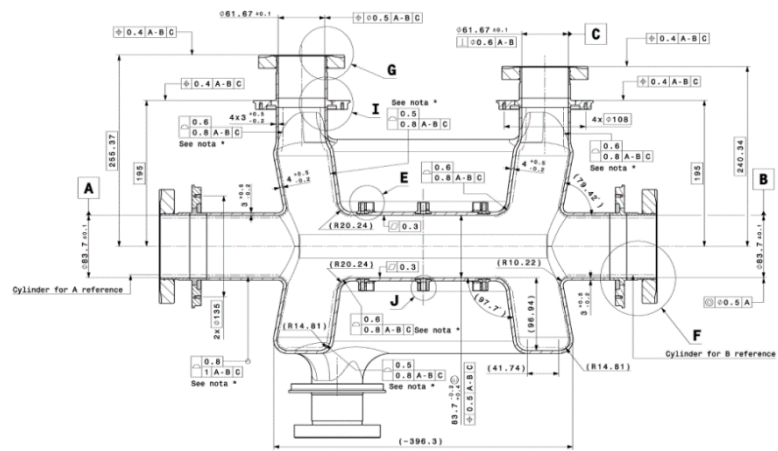

Figure 4-8: Dimensional plot with tolerances of the DQW cavity

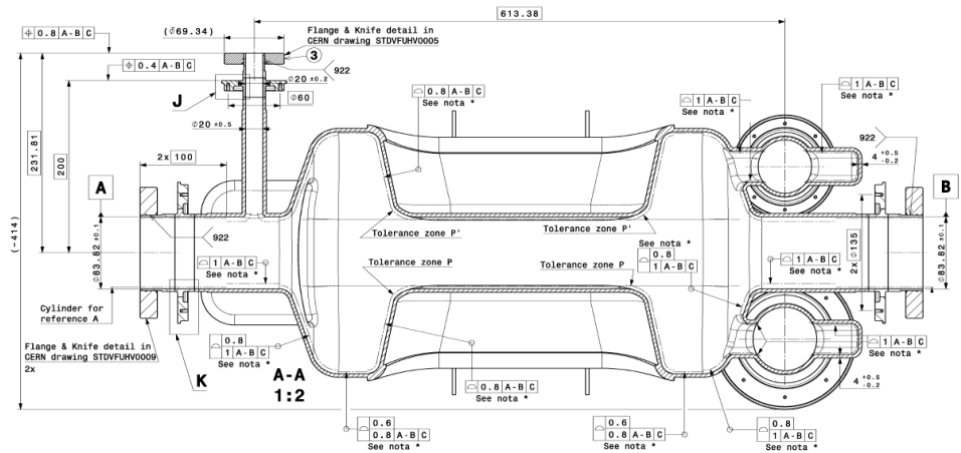

Figure 4-9: Dimensional plot with tolerances of the RFD cavity

#### 4.2.10.3  Helium vessel and dressed cavity unit

The helium tank will contain saturated superfluid helium at 2 K, cooling the cavity and allowing the extraction of the heat dissipated in the cavity and adjacent cold components. Superfluid helium is an excellent thermal conductor for small heat flux. Above a critical heat flux, the temperature increases drastically and eventually superfluidity is lost. The geometry of the helium tank has been determined to allow this maximum heat extraction while optimizing the quantity of the helium to be used.

Two choices of material have been studied for the helium tank: stainless steel and titanium. Titanium has the advantage of the same thermal contraction as niobium (in the order of 1.5 mm/m from ambient temperature to 2 K), while the thermal contraction of stainless steel is twice as large, leading to larger thermal stresses. The advantage of stainless steel is the manufacturability and thus the cost. However, for the unconventional geometries of the crab cavities, titanium grade 2 was chosen as the optimum material for the helium tank, allowing for rigid connection of cavity ports to the helium vessel.

The helium tank has a structural role, and its rigid connection to the cavity ports ensures optimum boundary conditions for the cavity during mechanical loading, in particular during maximum pressure loading and tuning. The helium tank geometry was chosen to limit the maximum stress on the cavity to tolerable values [12]. Figure 4-10 shows a qualitative stress distribution in the cavity wall during maximum pressure. The red colour indicates only small areas of high stress, which are tolerable. This distribution, as well as the maximum values, are directly influenced not only by the cavity geometry but also by the helium tank configuration.

A major concern for the mechanical design were the transitions from the helium tank to all of the adjacent components, in particular the main coupler, HOM couplers, and the flanges for connection to the beam pipes and helium pipes. All flange connections are stainless steel to stainless steel connections. Due to its proximity, the second beam pipe had to be integrated inside the helium vessel and consequently will be at 2 K; it is proposed to use a niobium beam pipe. A schematic view of the DQW and RFD cavities inside their helium tanks and equipped with the required ancillary equipment are shown in Figure 4-11 and Figure 4-12.

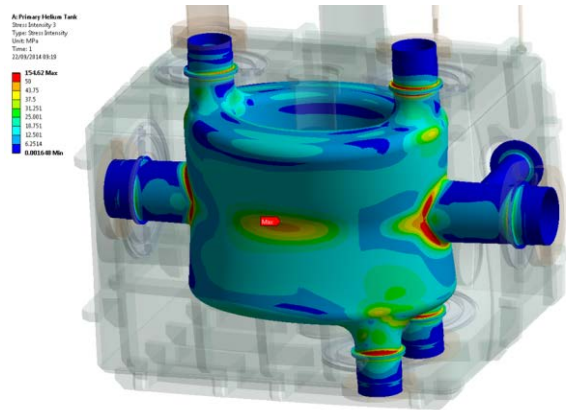

Figure 4-10: Mechanical stress induced by maximum pressure on the DQW cavity inside its helium tank. Red indicates regions with highest stress, which can be tolerated if confined to small areas.

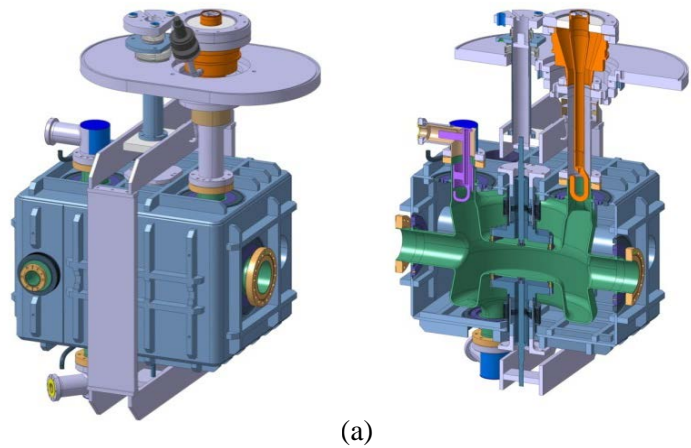

(a) (b)

Figure 4-11: (a) The DQW cavity inside its helium tank with the field probe port (front), beam port (right) and tuner frame around. (b) Sectional view of the DQW cavity inside its helium tank with the power coupler (top right, orange), HOM coupler (left, top and bottom), and tuner (centre, top, and bottom).

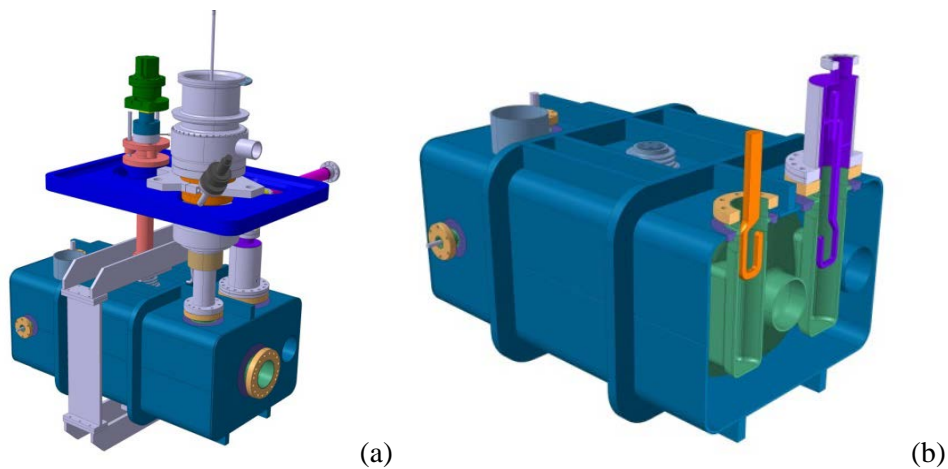

(a) (b)

Figure 4-12: (a) The RFD cavity inside its helium tank with the field probe port (centre left), beam port (centre right), tuner frame around helium vessel and tuner actuation (top centre). (b) Schematic sectional view of the RFD cavity inside its helium tank with the power coupler (orange) and HOM coupler (violet).

#### 4.2.10.4 Frequency tuning

The final resonance frequency of the cavity will depend on a number of fabrication and handling steps and cool-down (hundreds of kHz). A 'slow' mechanical tuning system is required to compensate for the uncertainties of the above steps by altering the cavity shape – this will dominate the tuner requirement. At 2 K, it must be possible to tune the cavity to $f_{res} = f_{operation} \pm \Delta f_{LFD}$, where $\Delta f_{LFD}$ denotes Lorentz force detuning occurring during cavity filling. The operating frequency can vary by an additional 60 kHz (cf. Section 4.2.18.5). Despite the large resulting tuning range (≈ ±200 kHz) the resolution of the tuner should allow at least four steps inside the cavity bandwidth (≈800 Hz); backlash and hysteresis must be small.

The tuning system, similar for both cavities (DQW and RFD), is shown in Figure 4-11 and Figure 4-12. It consists of an actuation system that is placed outside the cryomodule, and operated at room temperature and at atmospheric pressure, which makes it accessible and thus maintainable. The actuation system consists of a stepper motor, a harmonic gearbox, a roller screw, and linear guide bearings. The concept is based on a design developed and already in use at JLAB. The details of the prototype actuation system are shown in Figure 4-13. Since the cavity will be operated in continuous wave (CW) mode and frequency variations are expected to be small, active tuning with piezoelectric actuators may not be needed in the final design. A piezo is, however, foreseen for the first cavity tests to validate this assumption.

Actuation induces a relative movement between two titanium cylinders. The inner cylinder is directly connected to the top of the cavity, the outer cylinder to the bottom via a titanium frame. A symmetric deformation is thus applied simultaneously to the top and bottom of the cavity.

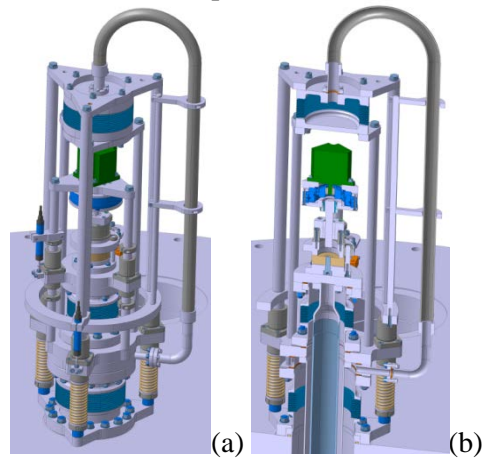

(a)   (b)

Figure 4-13: (a) Actuation system of the prototype tuning system for DQW and RFD cavities. (b) Cross-section.

The estimated mechanical resolution of the tuning at the connection to the cavity is estimated to be in the order of 0.1 μm, which is equivalent to a few tens of Hz for both cavities, allowing for at least 10 micro-steps inside the cavity RF frequency bandwidth.

Low frequency mechanical resonances (below 150 Hz) should be avoided to minimize cavity perturbation due to both helium pressure fluctuations $\mathcal{O}(1\text{ mbar})$ and external noise sources. Resonances above 150 Hz are considered to be benign. If fast-acting tuners (piezos) are deemed necessary, they should be able to compensate for deformations of ≤10–20 μm to reduce the RF power overhead (see Figure 4-14).

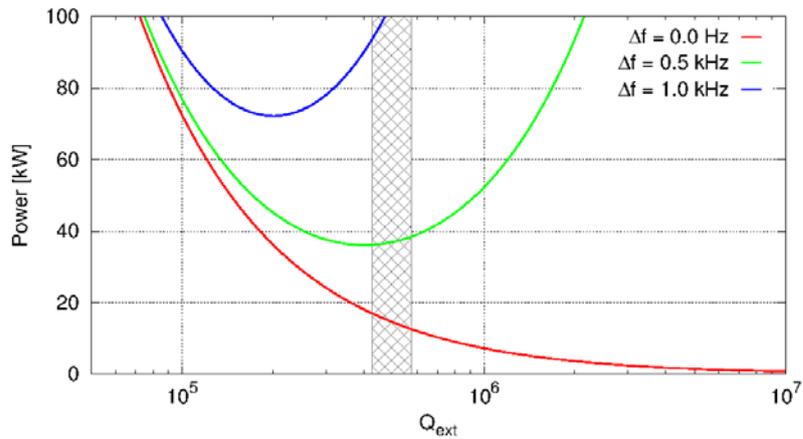

Figure 4-14: Forward power requires a function of $Q_{ext}$ for different detuning of the cavity. The cross-hatched area indicates the nominal range of $Q_{ext}$.

#### 4.2.10.5 Space, modularity, and the second vacuum chamber

Machine architecture and integration studies for the LHC led to the choice of housing two individual cavities in stand-alone cryomodules, individually connected to a cryogenic distribution line cryostat running in parallel with the main line. The nominal configuration will use a two-cavity cryomodule as a basic unit. As a consequence, a total of eight cold-to-warm transitions for the beam tube and four connections to the cryogenic distribution line are required for one side of an LHC interaction region (Figure 4-15).

The length of the cryomodule depends on the cavity type and, for the longest cavity, results in a total of 13.4 m for eight cavities (four cryomodules) per side of the LHC interaction region for both beams including gate valves from the interconnection plane, as shown in Figure 4-15. For each two-cavity module, two gate valves inside the cryomodule vacuum (see Figure 4-16) and two valves outside at ambient temperature are foreseen.

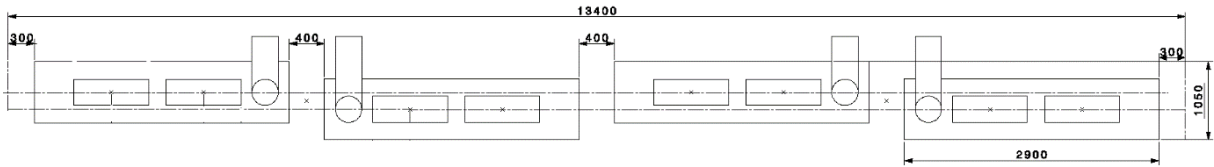

Figure 4-15: Cryomodule layout for one side of the interaction region in the LHC

A detailed view of the cryomodule containing two DQW and RFD cavities is illustrated in Figure 4-16. The fixed RF coaxial coupler, with a single ceramic window, providing 80 kW average power, is mounted onto the cavity via a ConFlat™ flange assembly equipped with a specific vacuum/RF seal designed at CERN and widely used elsewhere.

The RF coupler is mounted on the cavity in the clean-room, constraining the assembly of subsequent components of the cryomodule due to its size. The vacuum vessel was designed in three parts and uses a lateral assembly procedure for the cavity string inside the vessel [16]. This allows the possibility of cavity alignment with optical devices (laser trackers, for example) while making fine adjustments through the adjustable supports before closing the cryomodule lateral covers.

The cavity supporting concept uses the external conductor of the RF coupler as the main mechanical support of the dressed cavities. An additional supporting point to keep cavity alignment stability within requirements is obtained by the inter-cavity support. In the RFD cavity, the power coupler is transversely offset from the cavity axis, which requires additional vertical support, as shown in Figure 4-16(b).

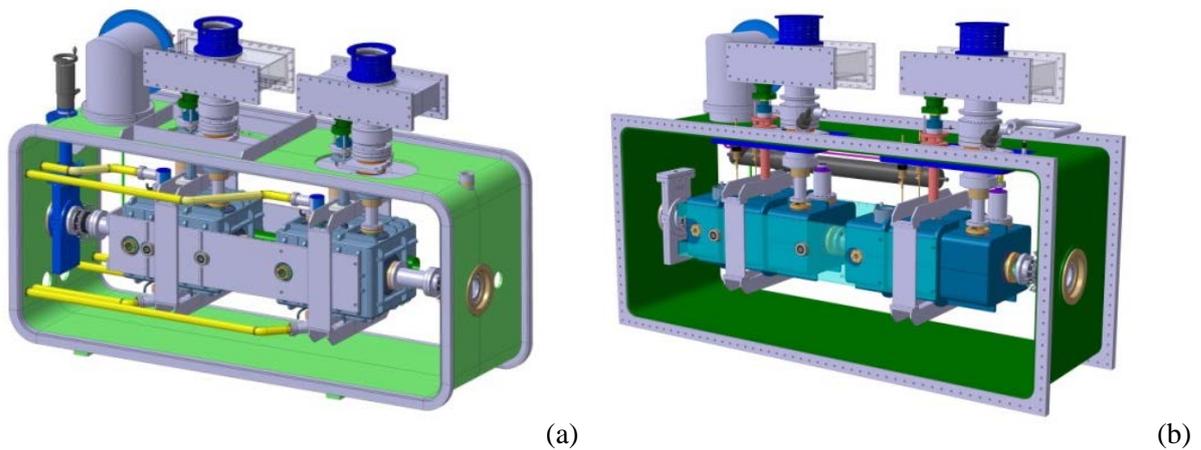

Figure 4-16: Cryomodules for (a) DQW cavity; (b) RFD cavity

For the LHC cryomodule, two options are considered. The baseline consists of four cryomodules per side per IP, each with two cavities similar to the SPS test prototype cryomodule. This would have the advantage of a topology similar to that having been tested in the SPS. The overall design would probably become simpler than for the SPS test prototype cryomodule. As a second option, a single eight-cavity cryomodule could be considered, optimized for LHC operation requiring less access and minimizing cold-to-warm transitions.

Currently only the SPS test cryomodule exists with full technical specifications. The cryomodules are designed to have a rectangular outer vacuum vessel with removable side panels such that the dressed cavities are side-loaded into the vessel [16]. All external connections except the beam pipes are on the top of the cryomodule. The cavities are supported by the power couplers. This allows easy access, as required for a prototype. This design requires several stiffening ribs to keep the stress within reasonable limits when placed under vacuum pressure and during cool-down. The designs for both cavity variants are kept as similar as possible.

#### 4.2.10.6  Magnetic and thermal shielding

Assuming a cavity geometric factor of $G \approx 100\ \Omega$, the additional surface resistance due to trapped flux $R_{mag}$ required to be below 1–2 n$\Omega$ to stay in the shadow of the total surface resistance specification of 10 n$\Omega$. To achieve this, magnetic shielding in the cryostat should reduce the external magnetic field on the outer surface of the cavity by a factor of at least 100 (reducing the earth's magnetic field to <1 µT).

The external warm magnetic shield is made of 3 mm thick mu-metal and will be directly attached to the vacuum vessel. Due to the large apertures in the shielding for couplers and beam pipes, this layer on its own is not sufficient to completely shield the earth's magnetic field to the required level with sufficient safety margin. Figure 4-17(b) shows the magnetic field amplitude inside a two-cavity cryomodule without an internal shield for an applied external shield of 60 µT in the longitudinal direction. To meet the magnetic field requirements a second shield is required close to the cavity. In order to reduce the size of the holes in the internal shield the cold magnetic shielding will be integrated inside the helium vessel, as presented in Figure 4-17(a). The internal shield is 1 mm thick and will be made from Cryoperm or Aperam Cryophy. Magnetization of both materials is adversely affected due by stress. Hence, degradation of the shielding material during assembly and handling should be carefully studied and monitored. Effects of weight and thermal stresses were modelled in ANSYS, as shown in Figure 4-17. The simulations indicate that while the maximum stress is 439 MPa in the titanium supports, the stress on the shield is kept to less than 150 MPa. It is possible that this may affect the magnetization locally, but the effect is comparable to that of a small hole in the shield. Simulation results from OPERA, assuming the worst case field orientation, show that the use of the proposed two-layer shielding solution to achieve magnetic fields well below 1 µT is feasible, as shown in Figure 4-17(b) [17].

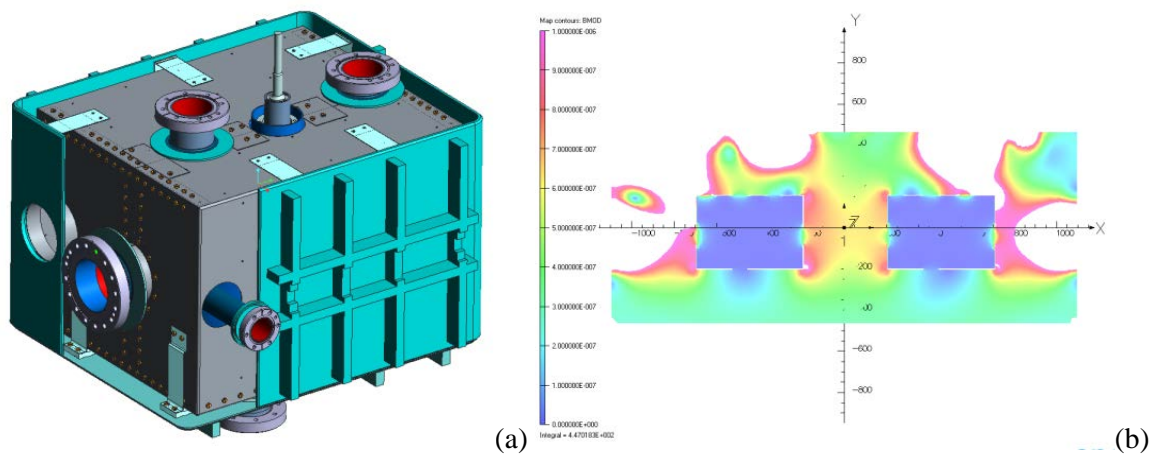

(a)                                                                (b)

Figure 4-17: (a) Cold magnetic shielding inside the helium vessel; (b), magnetic field amplitude inside the two-cavity CM without the second internal cold magnetic shield, scale 0 to 1 µT. An external field of 60 µT in the direction parallel to X (longitudinal) is used.

The thermal shield is made of rolled aluminium sheets. The shield is suspended from the vacuum vessel via flexible guides made from titanium alloy that also copes, through angular movements, with its thermal contractions. The absence of mechanical contact between the shield and the string of cavities eliminates the risk of interference with the alignment of the cavities induced by differential contractions and cooling transients. The cryomodule contains a single thermal shield, actively cooled in the LHC at about 50 K by a cryogenic cooling line containing pressurized helium gas. For the SPS tests, this active cooling will be done with pressurized liquid nitrogen. A 30-layer prefabricated Multi-Layer Insulation (MLI) blanket protects the thermal shield whereas a 10-layer blanket is mounted around each helium vessel.

4.2.11 RF powering and control architecture

The overall architecture and approximate volume of the RF infrastructure is shown schematically in Figure 4-18. Near P1 and P5, the existing caverns closest to the cavity (RR caverns) are approximately 80 m away, while requiring a large space in the tunnel to pass the RF transmission lines along this distance. Radiation concerns rule out the installation of highly sensitive RF electronics in those caverns. Therefore, two remaining options are under study.

- RF gallery near crab cavities. The longitudinal range would be approximately 155 m from IP1 and IP5 on either side in a gallery parallel to the LHC tunnel. Access to the gallery is required with RF on and field in the cavity, but without circulating beam.

- Surface option. The longitudinal range is similar to that given above, preferably above the crab cavity locations on the surface on either side of IP1 and IP5. To minimize the high power RF transmission line dimensions, the circulator and load are assumed to be in an extended tunnel alcove close to the cavity.

An independent powering system using LEP-type 400 MHz tetrodes (or an equivalent Inductive Output Tube, aka IOT) of 40–80 kW is assumed. Recent advances in solid-state technology could eventually lead to power sources in the required power range and may provide a cheaper and more robust platform. The tetrodes provide adequate power overhead in a compact footprint. This scheme would also allow for fast and independent control of the cavity set point voltage and phase to ensure accurate control of the closed orbit and the crossing angle in the multi-cavity scheme. Most importantly, fast control of the cavity fields will minimize the risk to the LHC during an abrupt failure of one of the cavities, ensuring machine protection before the beams can be safely extracted. For such fast and active feedback, a short overall loop delay between the RF system and the cavity is required [11].

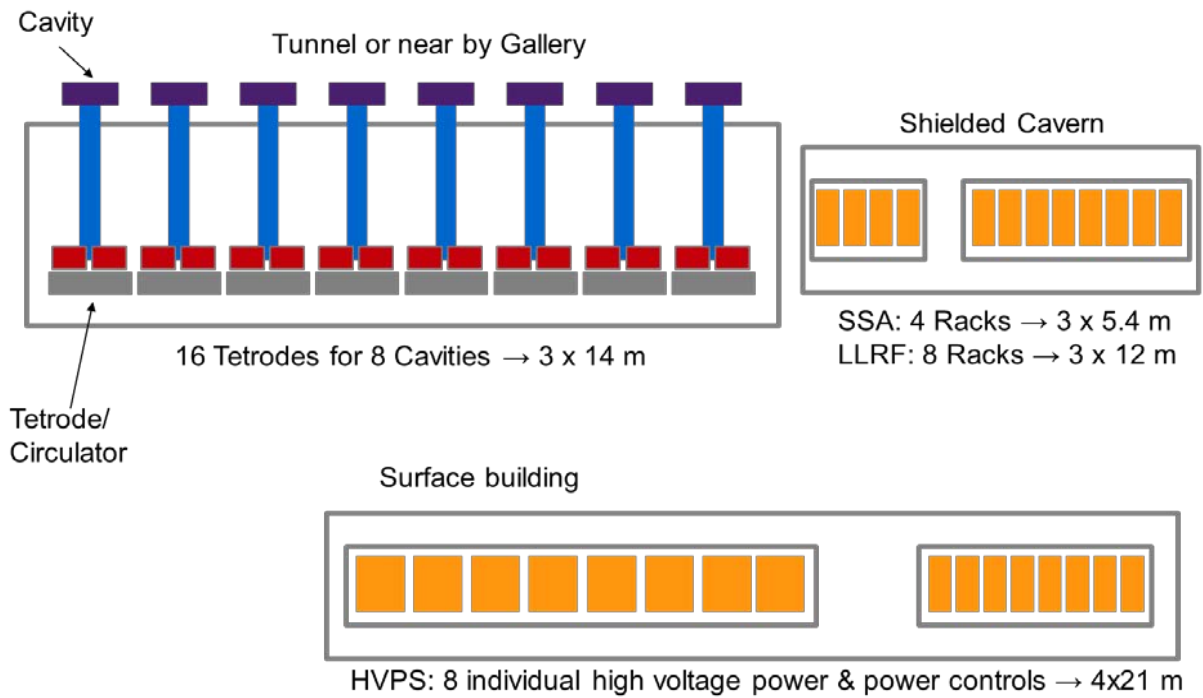

Figure 4-18: Schematic of the RF system layout in (a) the LHC tunnel; (b) the electronics racks in a shielded cavern close to the cavities; (c) the surface building. Note that these are only estimated values of space requirements.

To provide strong feedback, the low-level RF system requires the total loop delay to be approximately 1–2 μs. This includes the group delay from the driver, amplifier, circulator, and cable delays. Therefore, a distance of less than 100 m is desired for the separation between the amplifier, electronics, and the cavity in the tunnel. Such a short delay is already in place for the ACS main RF system in P4 (650 ns loop delay) with a service gallery running parallel to the tunnel.

The controls and driver electronics are required to be located in a radiation-minimized zone. Assuming two tetrode amplifiers per cavity to provide 80 kW and electronics racks required for drivers, PLC, LLRF, and fast interlocks for eight cavities per IP side, an area of approximately 100 $m^2$ is needed near the cavities. The high-voltage power supplies and the power controls would need an additional 85 $m^2$, which will nominally be placed on the surface. If all high- and low-power RF and controls are placed on the surface, an equivalent of 185 $m^2$ will be required there. The proximity of the circulator and RF loads to the cavity will allow for smaller RF transmission lines from the surface to the tunnel.

A total height of approximately 6 m is estimated for the high-power RF, controls, and services, distributed over two levels. This allows a minimum space of 3 m for the equipment racks, amplifiers, etc. while leaving 3 m height directly underneath cooling pumps, cabling, and services, see Figure 4-19. Alternatively, additional volume adjacent to the building to accommodate the pumps, ventilation and other required units for high power amplifiers can be envisioned to limit the height. All 'surface' buildings could actually be underground and would not occupy any surface area. The required electrical interfaces are specified in Ref. [18]. A study is ongoing to determine the feasibility of the civil engineering with minimal perturbation affecting the LHC [19].

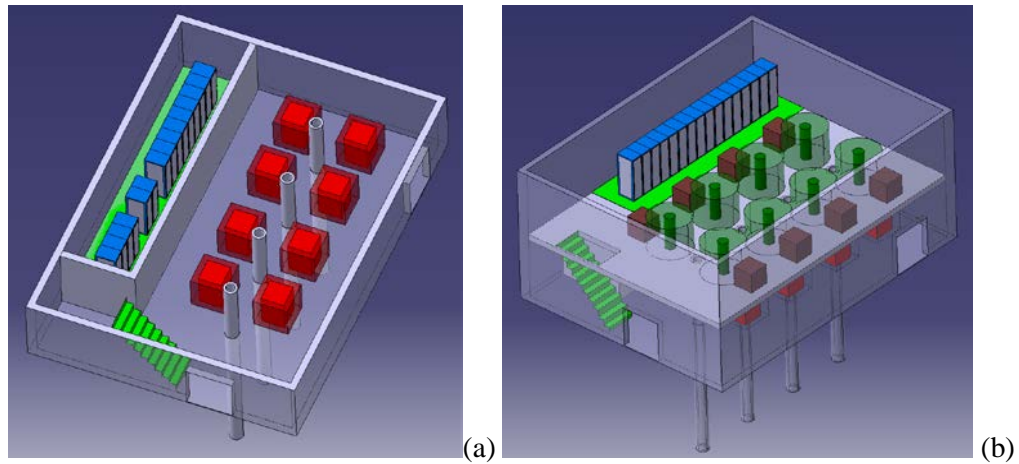

Figure 4-19: Preliminary sketch of a high-power RF, controls and LLRF layout in a surface building, distributed over two levels. (a) Lower level containing pumps and general services; (b) upper level containing RF amplifiers and equipment racks. Courtesy of C. Magnier and P. Fessia.

4.2.12 Low level RF architecture and operational scenarios

The RF control system, also commonly referred as the low level RF system (LLRF), includes several functionalities. First, a tuning control is required to keep the cavity resonant frequency on-tune with the beam during the crabbing operation. If required, the LLRF also has to ensure that the cavity is safely parked at an optimal detuned position during filling, ramping, and collisions without crabbing. This system also synchronizes the phase of the RF kicks with the exact passage of the bunches for both beams. The LLRF includes a regulation loop around the amplifier (to reduce the RF amplitude noise and phase noise in a band extending to a few tens of kHz), plus an RF feedback to control the cavity field precisely. The feedback loop consists of both a local loop around the cavity-amplifier and a global loop regulating the vector sum of voltages on the two sides of the interactions' region. The global loop will reduce beam perturbation following a single cavity trip, by quickly reducing the field in the companion cavities to track the uncontrolled voltage in the faulty cavity. The beam dump system has a three-turn (270 μs) response delay.

For each ring, the eight accelerating cavities (ACS) are driven from a single reference generated in a surface building above IP4. These two signals must be sent over phase-compensated links to IP1 (ATLAS) and IP5 (CMS). The eight crab cavities of a given ring at each IP are coupled with an 8-in, 8-out multi-cavity feedback (MFB). Figure 4-20 shows the proposed architecture.

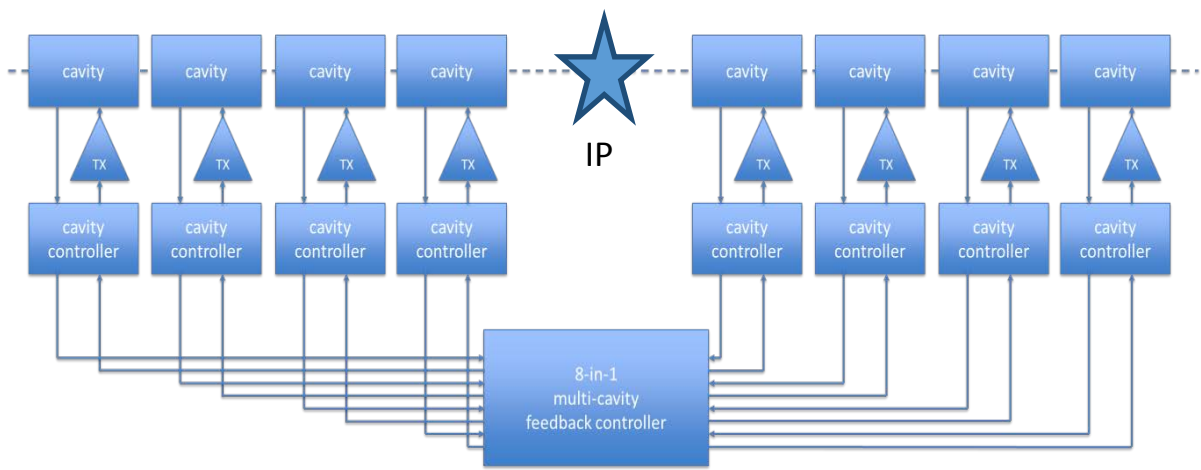

Figure 4-20: Proposed LLRF architecture for one ring at one IP

A central controller receives measurements from all relevant cavities on each ring and IP, and makes corrections to the drive of each individual TX. If the field starts changing in a cavity, the MFB will adjust the field in the other cavities on both sides of the IP, such that the orbit distortions remain local. As described in Section 4.2.5 on flat-top, counter-phasing is nulled while keeping the voltage set point small. The RF feedback keeps the cavity impedance small (beam stability) and compensates for beam loading as the cavity moves to resonance. The voltage set points are ramped to synchronously change the voltage in all crab cavities as desired. Any levelling scheme is possible. With a circulator between amplifier and cavity, the TX response is not affected by the cavity tune.

At present the spacing between LHC bunches within a batch is strictly constant along the ring. A large amount of RF power is used in the ACS system to fully compensate the transient beam loading caused by the 3 μs long abort gap and the smaller gaps required for the injection kicker ('half detuning'). This scheme cannot be extended into the HL-LHC era as it would require excessive RF power. The power required is minimized by optimally detuning the cavity ('full detuning') and adapting the cavity set-point phases bunch by bunch. It results in bunch arrival time modulation of up to ±42 ps [20]. This may be acceptable given the 1 ns bunch length. There is no effect on the luminosity as the modulation is identical in both beams, only the vertex position is modulated around the nominal vertex by a maximum of 1 μm over one turn. The bunch-to-bunch variation within a batch is at least an order of magnitude smaller. If not, the LLRF must synchronize the bunch-by-bunch crabbing field with the actual phase modulation.

### 4.2.13 Cavity failure scenarios

Crab cavity failures can lead to a fast voltage and/or phase change with a fast time constant. This can lead to large, global head–tail oscillations, or coherent betatron oscillations with a change in transverse beam trajectories of 1.7 $\sigma$ for a single cavity failure; the effect is cumulative with the number of failing cavities. These failures can be broadly classified into two categories.

- Fast failures, single or few turns. For example, a sudden cavity quench or breakdown.
- Slow failures, several tens of turns or greater (caused by vacuum degradation, voltage and phase drifts, or similar).

Due to the relatively high quality factor in the superconducting cavity, the stored energy inside the cavity can typically only be extracted with a time constant determined by $Q_L$, which results from the strong coupling to the cavity via the power coupler. The stored energy will decay with a time constant $\tau = 2Q_L/\omega_0$. For $Q_L = 5 \times 10^5$, the time constant is approximately 400 μs. The three turn delay time (267 μs) for a beam dump trigger is an important consideration during a RF source failure, where the cavity field decays to roughly half its value before the beam can be safely aborted. In the case of a quench, the time constant of field decay may be dominated by the quench dynamics rather than $Q_L$. The situation is similar due to strong and sudden electron loading due to multipacting or other phenomena.

The cavity quench mechanism described above and measurements from KEKB crab cavities [21] indicate that typically a quench is a slow thermal process (typically of the order of several milliseconds). Once the temperature of a sufficiently large area exceeds the critical temperature of niobium, the quench can propagate very quickly to completely quench or cause RF breakdown. However, any change in cavity quality factor well before reaching a critical temperature limit could be easily detected from the requested forward power (fast) or changes in the cavity temperature bath (slow). An interlock on the forward power, except due to induced orbit excursion, can cut the RF to slow down or stop quench propagation. A beam abort, if required, can be triggered simultaneously (a few μs) for machine protection.

### 4.2.14 Failure scenario mitigation

The choice of low operating temperature (2 K) and moderate surface field levels allow operation with an ample margin over quench temperature and field limits. The significantly better thermal conductivity of superfluid helium should also improve the thermal performance and stability of the cavity. Additional measures in the cryomodule design are being considered to dimension the helium enclosures with sufficient margin for heat flux. The cavity thermal and RF stability will be thoroughly tested in the SM18 test facility and during the SPS beam tests.

To minimize the perturbation on the beam during a cavity failure, the MFB will adjust the field in the other cavities on both sides of the IP, such that the orbit distortion remains local. Figure 4-21 shows the cavity control of two cavities across the IP with one cavity failure and the RF controller to adjust the second cavity to follow. The rapid change in field will also result in a detuning of the cavity; however, the mechanical tuning system is unable to adjust the tune within 400 μs. Since a rapid breakdown of a failed cavity may become unpredictable, it is probably safest to ramp down the cavities synchronously. However, small and slow changes in one of the cavities can be adjusted for without aborting the beam.

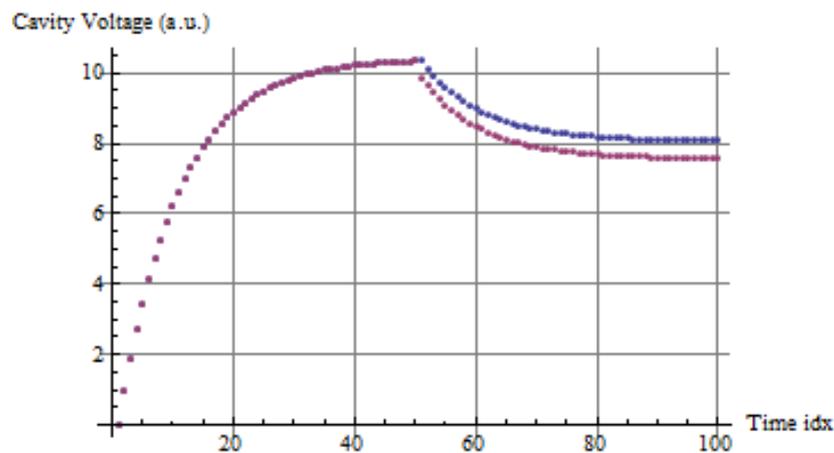

Figure 4-21: Voltage response with strongly coupled cavities across the IP as a function of time [μs]. At 50 μs, one cavity trips (red trace) and the other one is forced by the RF controller to follow (blue trace).

The cavities can be equipped with a fast tuning system such as a piezo mechanism. If the speed of such tuning devices is sufficient, it could compensate for Lorentz force detuning during transients and thus keep the tune within the bandwidth of the feedback system.

An additional mitigation to avoid large beam losses (and hence deposited energy) in the case of single or multiple cavity failure is a robust measurement and interlocking of the tail population and eventual head–tail oscillations. This could be achieved with new equipment such as a hollow electron lens for cleaning of the bunch tails and interlocking with improved diagnostics like fast head–tail monitors and/or fast beam loss monitors (e.g. Diamond monitors).

4.2.15 Heat loads and cryogenics

The cavities are housed in individual titanium helium tanks connected by a 100 mm diameter two-phase He pipe placed above the cavities. This pipe ensures that liquid is fed to the cavities by gravity, and is also used as a pumping line for gaseous helium. A saturated helium bath maintains the cavities operating temperature at 2 K. Liquid helium is supplied to the two-phase pipe through a capillary. It is proposed to fill from one single point at one extremity of the two-phase pipe, and control just the He level in the phase separator. The static plus dynamic heat loads are expected to be approximately 30 W to the 2 K bath for a two-cavity module. The cryogenic limits in the LHC are not precisely known at this time. However, the 15 W per cavity heat load at 2 K is small compared to the LHC heat load capacity; the total heat load of the LHC crab cavity systems is estimated at 0.5 kW at 2 K.

4.2.16 Vacuum system

The crab cavity system has three independent types of vacuum systems: the cavity vacuum, the adjacent beam pipe, and the cryostat. The two-cavity common vacuum is pumped at room temperature with two ion pumps mounted at each end of the modules. However, at 2 K, the cryogenic pumping of the cavity walls is the dominating feature, with a pumping speed of 10 000 L/s. The background pressure without RF is expected to much better than $10^{-10}$ mbar and likely limited by the measurement devices such as Penning gauges. Pressure signals provided for RF control are a hardware interlock from the ion pumps to cut the high voltage and readout from the Penning gauges, one per coupler, to limit the RF power. The cavity vacuum can be isolated by two all-metal valves at the ends of each module, to maintain vacuum during transport and installation.

The second beam pipe for the counter-rotating beam has to pass through the cavity helium vessel due to its proximity. It is planned that this will be made of niobium and will remain superconducting at close to 2 K to preserve the same surface conditions as in the cavity. The use of carbon coating in the warm regions near the crab cavities to reduce the pressure and to avoid electron cloud effects is currently not considered; this would risk contamination of the cavities.

The insulation vacuum is less demanding in terms of pressure, the modules being pumped to $10^{-5}$ mbar before being cooled down. When cold, the insulation vacuum also benefits from the cryogenic pumping of the cold surfaces and the operating pressure will decrease to $10^{-7}$ mbar. Turbo molecular pumps are used and pressures are measured using Penning gauges. Intense gamma radiation could be produced during cavity RF conditioning and operation of high fields.

4.2.17 Interlocks for machine protection

Due to the immense stored energy (>700 MJ), the transient behaviour of the crab cavities is of concern. The crab cavity system will be equipped with several levels of interlocks both for machine protection and to protect the RF system itself. Slow and fast interlocks, including specific RF interlocks (reflected power, signal level, arc detection, etc.) will ensure safe operation under all conditions and cope with transients; the interlock system will be fully embedded in the overall machine interlock system. All RF systems, including amplifiers, circulators, and loads are designed to withstand full reflection in the case of a malfunction in the RF chain.

4.2.18 SM18 and SPS beam tests

The addition of crab cavities to the LHC should ensure robust functioning through the entire sequence of the LHC physics cycle. Since crab cavities of this type have yet to be realized and used with hadrons, beam tests with a prototype two-cavity cryomodule are a prerequisite to identifying potential risks from the technology to safe and reliable operation of the LHC. Therefore, an essential milestone for a crab cavity in the SPS is to demonstrate machine protection and cavity transparency. All RF manipulations and cavity-beam interactions will first be validated and commissioned in the SPS with the prototype module. The beam tests are planned as machine development studies during the run 2017–2018. Successful validation of the crab cavities in the SPS is a prerequisite for installation in the LHC.

**4.2.18.1 Tests before installation**

The two cavity cryomodule will be assembled in SM18 with the cold masses (cavities, helium vessels, HOMs, and tuner) provided by the LARP collaboration. The cold masses are tested and qualified in the US to their final specification and delivered under vacuum to the SM18 facility. Only the assembly of the power coupler in SM18 is foreseen, due to the risk associated with damage during transport.

The assembled SPS cryomodule in its final assembly, along with all of the other major components, will be tested for vacuum integrity, RF performance, and operational reliability in the SM18 horizontal bunker to their nominal specifications prior to installation in the SPS. The RF control and interlock system will also be validated in SM18.

### 4.2.18.2 SPS environment

The SPS ring is equipped in LSS4 (currently used for the COLDEX experiment) with a special bypass (Y-chamber) with mechanical bellows that allows for horizontal displacement. This allows for a test module to be moved out of the beam line during regular operation of the SPS and only moved into the beam line during the periods dedicated to studying the crab cavity test module with beam. This setup is essential both due to aperture limitations of the crab cavities and the risk associated with leaving the cavities in the beam line with different modes of operation in the SPS. The cryomodule is placed on a movable table that can be moved sideways by 510 mm (see Figure 4-22). A special working group (crab cavity technical coordination) has been set up to follow up the various integration issues including the RF and cryogenic systems.

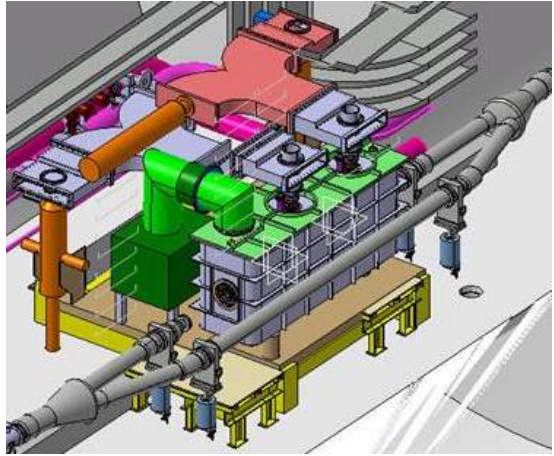

Figure 4-22: 3D integration of the cryomodule, RF assembly, and the cryogenics in the SPS

The relevant cryomodule envelope dimensions for the SPS tests is given in Table 4-3. In general, the SPS constraints are tighter than those of the LHC.

Table 4-3: Cryomodule envelope dimensions

| Description | Distance [mm] |
|---|---|
| Cryomodule length (gate valve to gate valve) | 3000 |
| Horizontal distance cavity axis to inner edge of cryomodule volume | 420 |
| Vertical distance, floor to cavity axis | 1200 |
| Maximum height above cavity axis | 1200 |
| Inner diameter of cavity beam pipe | 84 |
| Horizontal distance cavity axis to bypass beam pipe axis | 510 |

### 4.2.18.3 SPS RF system and operation

Specially designed WR2300 will feed the RF power from the tetrode amplifiers to the respective cavity (see Figure 4-23). Placement of the amplifiers on the movable table will depend on the full integration of the cryomodule, transmission lines, and circulator. An LHC-type circulator, although over-dimensioned, is preferred for reasons of maintenance and spares policy. A 3D integration of the cryomodule and the RF assembly in the LSS4 region is shown in Figure 4-22.

In the LHC, the four cavities per IP side are always powered on tune, initially with a small voltage (10%–15% of the nominal) and counter-phased with active feedback to guarantee maximum beam stability during the entire cycle. Therefore, beam injection with counter-phased cavities with low voltage requires testing in the SPS. Other issues related to beam loading and transient effects with and without RF feedback and slow orbit control will be studied to evaluate the stability and tolerances required from the feedback systems. Induced RF trips and their effects on the beam will be studied in detail to guarantee machine protection and to devise appropriate interlocks. Long-term effects with crab cavities on coasting beams at various energies will also be tested.

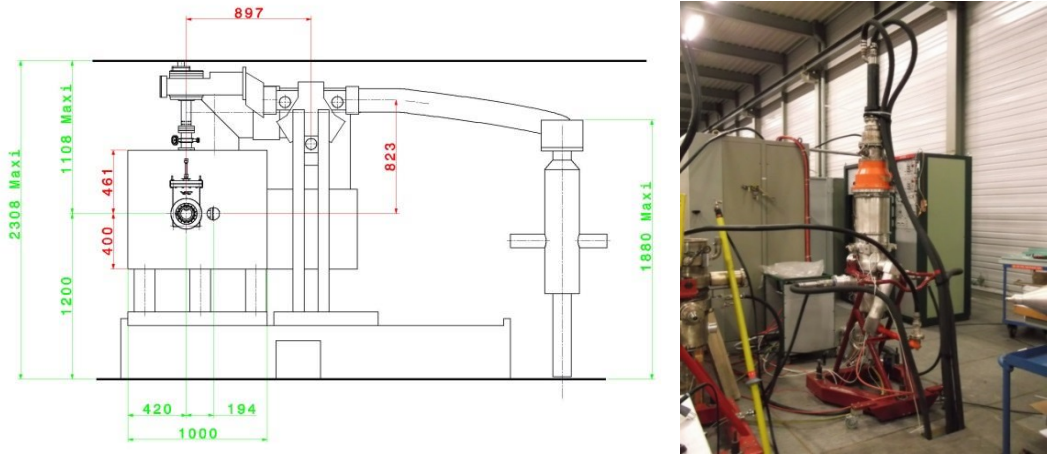

Figure 4-23: (a) Cryomodule and RF system layout in the LSS44 cavern; (b) a 400 MHz tetrode amplifier under test.

#### 4.2.18.4 SPS cryogenics requirements

The SPS-LSS4 region is currently equipped with the TCF20 cryogenic box. It was originally planned to upgrade the existing cold box to deliver 2 K helium for the operation of the crab cavities [15]. Due to the very limited capacity available in the SPS TCF20 refrigerator, the total losses (static and dynamic) are limited to a maximum of 25 W (at 2 K). The heat load of the two-cavity cryomodule, however conservative, is at least 15% higher than the TCF20 capacity. A replacement for the TCF20, increasing the capacity to approximately 40 W, is planned to be put in place for the SPS tests [15]. Despite the plan to increase the capacity, a strong effort is being mae to minimize the heat load of all cavity elements and cryomodule interfaces to ensure a successful beam test in the SPS prior to any installation in the SPS.

#### 4.2.18.5 SPS tuning requirements

For beam tests in the SPS a slow mechanical tuner system is required to bring the cavity on resonance in the energy range of the SPS (0–60 kHz). In addition the tuner must allow detuning of the cavity to its parking position, and it has to be precise enough to work together with the RF feedback. Table 4-4 summarizes the potential energies at which the SPS can be operated for crab cavity tests and their corresponding RF frequencies compared with that of LHC operation.

Table 4-4: Detuning ranges for the LHC and SPS

| Parameter | Unit | LHC | SPS | | |
|---|---|---|---|---|---|
| Energy | [GeV] | 450–7000 | 120 | 270 | 450 |
| Frequency | [MHz] | 400.79 | 400.73 | 400.78 | 400.79 |
| $\Delta F_0$ | [kHz] | 0 | −58.2 | 12.2 | −2.4 |
| Bandwidth | [kHz] | 0.4–4 | 0.4–4 | 0.4–4 | 0.4–4 |
| Detuning | [Hz] | ±5.5 | ±21.7 | | |

The detuning required to tune the cavity in its parking position (between betatron lines) is approximately ±21.7 kHz in the SPS. The detuning requires a resolution of at least one-quarter of the final cavity bandwidth due to available power limits. Additional studies have to be carried out to verify if a tuning speed higher than that possible with the mechanical tuner is required if limitations arise from feedback and/or orbit control.

#### 4.2.18.6 SPS test objectives

The test programme objectives in the SPS are given below.

- Demonstration of cavity deflecting field with proton beam including injection, energy ramp, and coast at energies ranging from 26–450 GeV.
- Verification and control of cavity field (amplitude and phase), frequency, tuning sensitivity, input coupling, power overhead, and HOM signals. Establish and test operational cycle with crab cavities.
- Demonstrate the possibility to operate without crab cavity action (make them invisible) by both counter-phasing the two cavities or by appropriate detuning (to parking position) at energies ranging from 26–450 GeV.
- Measurements of beam orbit centring, crab dispersive orbit, and bunch rotation with available instrumentation such as BPMs and head–tail monitors.
- Demonstrate MFB operation.
- Demonstrate non-correlated operation of two cavities in a common cryomodule – trigger quench in one cavity without inducing quench in the other.
- Define and implement interlock hierarchy. Verification of machine protection aspects and functioning of slow and fast interlocks.
- Test HOM coupler operation with high beam currents, different filling schemes, and associated power levels. Measurement of impedance and instability thresholds for nominal mode and HOMs.
- Measure emittance growth induced by the crab cavities as far as possible.

#### 4.2.18.7 Outline of an SPS test programme

- Initial RF commissioning with the cryomodule in out-of-beam position (no dedicated MD required).
- RF commissioning with low-intensity beam, single bunch to a few bunches. Establish the proper RF parameters, including cavity tune, operating frequency, amplitude, and phase. Verify crab cavity active and invisible.
- High intensity single bunches to trains of bunches to investigate the effect of cavity performance, impedance, and machine protection; and characterize the transient behavior of the crab cavity system as a function of beam current. Verify cavity stability over many hours (as relevant for LHC physics fill).
- Long-term behavior of coasting beams in the SPS with relatively low intensity to study the effects of emittance growth and possibly non-linear effects such as RF multipoles

### 4.3 Harmonic systems

The harmonic cavity systems are presently not part of the HL-LHC baseline. Two categories of harmonic systems are proposed [20], [22].

- A higher harmonic (800 MHz) system can be used either for changing the bunch profile (in bunch lengthening mode (BL)) or for increasing the synchrotron frequency spread (in BL or bunch shortening mode (BS)). Depending on the mode of operation, this RF system can help to reduce the beam-induced heating, effect intra-beam scattering, improve longitudinal beam stability, and in some scenarios increase or level luminosity.
- A sub-harmonic (200 MHz) system could either completely replace the existing main RF system or work jointly with the 400 MHz RF system, which in this case will act as the second harmonic. The lower harmonic RF system will improve the capture efficiency for longer SPS bunches with very high intensity. The benefits of the combined 200 MHz and 400 MHz system are similar to the above double-harmonic system but with the primary aim of luminosity improvement.

For the higher harmonic system (800 MHz), a maximum of 8 MV longitudinal voltage can be provided from approximately four to ten cavities depending on the mode of operation [20]. Relevant RF parameters are listed in Table 4-5. This is a maximum of 300 kW input power assumed to be feasible [23]. The BS mode, with the full-detuning scheme in the fundamental 400 MHz cavities, requires significantly lower RF power. Therefore, a four-cavity system is more than adequate to provide the required 8 MV with a maximum power of 300 kW per cavity. In the BL mode, the required RF power at 1 MV already exceeds 300 kW. Therefore, approximately 10 cavities are needed to provide for the 8 MV required to stay below the RF power limit.

Table 4-5: Relevant RF parameters for 800MHz RF cavities

| Characteristics | Units | Value |
|---|---|---|
| Resonance frequency | [MHz] | 801.58 |
| Total accelerating voltage | [MV] | 8 |
| Number of Cavities | | 4 (BS) to ≈10 (BL) |
| Residual resistance $R_s$ | [nΩ] | ≈250 |
| $R/Q$ | [Ω] | ≈45 |
| $Q_0$ | | $\geq 1 \times 10^9$ |
| $Q_{ext}$ | - | N/A |
| RF power per cavity | [kW] | $10^5$ (BS), ≈$10^4$ (BL) |
| Operating temperature | [K] | 4.5 |

Replacing the existing acceleration system with a sub-harmonic system (200 MHz) will require a minimum of 3 MV of longitudinal voltage to capture, accelerate, and store the HL-LHC beams [24]. This can be provided from two to four compact quarter wave cavities, where the number of cavities also depends on power requirements and technology constraints [20]. Some relevant RF parameters for the 200 MHz cavities are listed in Table 4-6.

Table 4-6: Relevant RF parameters for 200 MHz RF cavities

| Characteristics | Units | Value |
|---|---|---|
| Resonance frequency | [MHz] | 200.4 |
| Total accelerating voltage | [MV] | 3–6 |
| Residual resistance $R_s$ | [nΩ] | ≤10 |
| $R/Q$ | [Ω] | ≈50 |
| $Q_0$ | | $\geq 1 \times 10^{10}$ |
| $Q_{ext}$ | - | N/A |
| RF power (assumed) | [kW] | 500 |
| Operating temperature | [K] | 4.5 |

The total static and dynamic heat load for either system has to be evaluated in detail during the engineering phase of the cryomodule. A cryogenic sectorization of two cavities per cryomodule is assumed for modularity, maintenance, and reliability. For 200 MHz, a preliminary estimate of 120 W for a two-cavity module at 4.5 K can be assumed where each cavity operates at 3 MV. The cavity technology (bulk niobium or Nb-coated copper cavities) can play a role in the final quality factor of the cavity and hence the heat loads at the operating gradient.

For the 800 MHz system, the frequency dependence of the surface resistance gives 250 nΩ leading to approximately 50 W at 4.5 K due to dynamic RF losses. This is only 1 W at 2 K. A geometric factor of 230 Ω and a cavity voltage of 2 MV are assumed. Therefore, it is preferable to operate the 800 MHz system at 2 K to both take advantage of the lower surface resistance and superior properties of superfluid helium. Assuming a baseline of 4.5 K, approximately 200 W can be assumed as an upper limit for a two-cavity module.

### 4.4 Transverse damper (ADT) upgrade

The LHC requires a transverse feedback to damp injection oscillations and provide stability for impedance-driven transverse instabilities, thus guaranteeing preservation of beam intensity and emittance [25]. The existing coupled bunch feedback system ADT, installed in P4 of the LHC, was fully commissioned in 2010 [26]. It damps transverse instabilities within a bandwidth of 20 MHz, correcting the oscillations of the centre of gravity of the individual bunches about their orbit.

For the upgrade of the ADT system, three possible routes have been identified in the past [27]: increase of kick strength, reduction of noise, and increase of bandwidth. A space reservation of approximately 5 m on each side was made in the original design of the LHC to install more kickers adjacent to the existing ADTs in P4 [28].

Following the experience of the LHC Run 1, priority was given to an upgrade of the pick-up and signal processing systems aimed at reducing the noise floor, one of the options already foreseen in 2006. This includes new electronics and a doubling of the number of pick-ups, from two to four per beam and plane, with commissioning foreseen after LS1 [29]. The experience of Run 1 has also shown that an increase in kick strength may not be required, as injection errors are on average more than a factor of 4 smaller than originally assumed, and fast damping times of less than 20 turns for the injection errors can be achieved with the existing system. During Run 1, improvements in ADT signal processing were tested to address single bunch oscillations observed, which were different in nature from the coupled bunch oscillations typically driven by the resistive wall impedance, which falls off in frequency towards 20 MHz. The nature of some of these observed instabilities was not entirely unravelled during Run 1, and new diagnostics such as the multiband instability monitor (MIM) were added to the LHC to better characterize these instabilities [30]. Improvements in ADT signal processing tested during Run 1 permit running with a flat frequency response of up to approximately 20 MHz [31]. Beyond 20 MHz, the kicker and ADT power amplifier system cannot be used.

For the HL-LHC, a transverse kicker system with a larger bandwidth than 20 MHz would be an asset in view of the high bunch intensity.

In the SPS a development was started in 2008 to design a high bandwidth transverse feedback system [32, 33] that aims at damping intra-bunch motion. The system consists of pick-ups, kickers, power amplifiers, and signal processing. Both slot-line and strip-line kickers are studied for this system; slot-line kickers could offer broadband response of up to 1.2 GHz [34]. The system can also be used to damp and attenuate intra-bunch motion caused by the electron cloud, impedance-driven instabilities, or other perturbations driven, for example, by beam–beam effects or crab cavities. With its generic approach the results of the SPS development are relevant to other accelerators, including the LHC and other large colliders.

At 1 GHz four slotlines, as developed for the SPS LIU project [34], with 2 kW amplifier power per coupling port can develop a transverse voltage of 37 kV. Consequently the beam can be kicked by 82 nrad at 450 GeV and 5 nrad at 7 TeV. Assuming a reference beta function value of 180 m, at pick-up and kicker, the kick strength corresponds to removing 15 µm of oscillation at 450 GeV and 0.9 µm at 7 TeV, within a single turn. As a feedback system with a gain corresponding to 500 turns damping time, saturation would occur at 3.7 mm at 450 GeV and 225 µm at 7 TeV.